\documentstyle[sprocl,epsfig]{article}

\newcommand{\re}{\mathop{\rm Re}\nolimits}
\newcommand{\im}{\mathop{\rm Im}\nolimits}
\newcommand{\alt}{\,\rlap{\lower 3.5 pt \hbox{$\mathchar \sim$}} \raise 1pt
 \hbox {$<$}\,}

\newcommand{\be}{\begin{equation}}
\newcommand{\ee}{\end{equation}}
\newcommand{\bea}{\begin{eqnarray}}
\newcommand{\eea}{\end{eqnarray}}
\newcommand{\bm}{\boldmath}
\newcommand{\ubm}{\unboldmath}
\newcommand{\smallz}{{\scriptscriptstyle Z}} 
\newcommand{\smallw}{{\scriptscriptstyle W}} %

\newcommand{\fr}{\frac}

\newcommand{\mz}{M_\smallz}
\newcommand{\mw}{M_\smallw}

\def \mev  {\mbox{ MeV}}

\def \psl  {p \kern-.45em{/}}
\def \qsl  {q \kern-.45em{/}}
\def \qslov {\overline{q \kern-.45em{/}}}
\def \pslov {\overline{p \kern-.45em{/}}}

\def \sov  {\overline{s}}
\def \mov  {\overline{m}}

\def \lsim {\raisebox{-.7ex}{$\stackrel{\textstyle <}{\sim}\,$}}

\def \xiw   {\xi_\smallw}
\def \xiz   {\xi_\smallz}
\def \xig  {\xi_\gamma}

\def \LM   {\ln \! \left(\frac{M^2-s}{M^2} \right)}

\def \LS   {\ln \!\left(\frac{\sov-s}{\sov} \right)}
\def \Agamma {A^{\gamma}}

\begin{document}


\title{\vskip-2.cm{\baselineskip14pt
\centerline{\normalsize\hfill MPI/PhT/98--88}
\centerline{\normalsize\hfill hep--ph/9811467}
\centerline{\normalsize\hfill November 1998}
}
\vskip.5cm
RECENT DEVELOPMENTS IN THE STUDY OF UNSTABLE PARTICLE
PROCESSES\footnote{To appear in the
{\it Proceedings of the IVth International Symposium
on Radiative Corrections (RADCOR~98)},
Barcelona, Spain, 8--12 September 1998, edited by J. Sol\`a.}
}

\author{ALBERTO SIRLIN\footnote{Permanent address: Department of Physics, New 
York University, 4 Washington Place, New York, NY 10003, USA}}

\address{Max-Planck-Institut f\"ur Physik (Werner-Heisenberg-Institut),\\
F\"ohringer Ring 6, 80805 Munich, Germany\\
E-mail: alberto.sirlin@nyu.edu}

\maketitle\abstracts{Developments in the analysis of $W$ and quark propagators
in the resonance region, and recent considerations concerning the mass and 
width of the Higgs boson are reviewed.
Particular emphasis is placed on the instability of these fundamental 
particles, and on related issues of gauge dependence.}

\section{Introduction}

We discuss two subjects:
\begin{enumerate}
\item Radiative Corrections to $W$ and Quark Propagators in the
Resonance Region and Related Problems.
\item On the Mass and Width of the Higgs Boson.
\end{enumerate}
We recall the conventional definitions for the mass and width of unstable
vector bosons:
\begin{equation}
M^2=M_0^2+\re A(M^2),\qquad
M\Gamma=-\frac{\im A(M^2)}{1-\re A^\prime(M^2)},
\label{eq:oms}
\end{equation}
where $A(s)$ is the transverse self-energy. More fundamental definitions are
based on the complex-valued position of the propagator's pole:{}\cite{var}
\begin{equation}
\bar s=M_0^2+A(\bar s),\qquad
\bar s=m_2^2-im_2\Gamma_2.
\label{eq:pol}
\end{equation}
We may identify the mass and width with $m_2$ and $\Gamma_2$.
Taking the real and imaginary parts of Eq.~(\ref{eq:pol}), we find
\begin{equation}
m_2^2=M_0^2+\re A(\bar s),\qquad
m_2\Gamma_2=-\im A(\bar s),
\label{eq:mgp}
\end{equation}
so that $\re A(\bar s)$ plays the r\^ole of mass counterterm. In terms of
$m_2$ and
$\Gamma_2$ the Breit-Wigner (BW) resonance amplitude is proportional to
$(s-m_2^2+im_2\Gamma_2)^{-1}$. As $\bar s$ is the
position of a singularity in the analytically continued $S$-matrix, it has
the important property of being gauge invariant.

If $\Gamma_2/m_2={\cal O}(g^2)$ is small, one may expand Eq.~(\ref{eq:mgp}) in
powers of $m_2\Gamma_2$ about $m_2^2$. One
readily finds that the result agrees with Eq.~(\ref{eq:oms}) in
next-to-leading order (NLO), but not beyond.
Given $m_2$ and $\Gamma_2$, other definitions are possible.
For example, in the $Z^0$ case,
\begin{equation}
m_1=\sqrt{m_2^2+\Gamma_2^2},\qquad
\Gamma_1=\frac{m_1}{m_2}\Gamma_2
\label{eq:lep}
\end{equation}
lead, to very good accuracy, to an $s$-dependent BW amplitude. An important
consequence is that $m_1$ and $\Gamma_1$ can be identified with the $Z^0$ mass
and width measured at LEP.
A third frequently employed parametrization is
\begin{equation}
\bar s=\left(m_3-i\frac{\Gamma_3}{2}\right)^2.
\label{eq:p3}
\end{equation}

Theoretical arguments led to the conclusion that, in the $Z^0$ case, $M$ is
gauge dependent in ${\cal O}(g^4)$ and higher.\cite{Si91a,Si91b}
This has been confirmed by an analysis of the $Z^0$ resonant propagator in
general $R_\xi$ gauge.\cite{PaSi96,rin} The
gauge dependence in ${\cal O}(g^4)$ is small ($\alt2$~MeV), but it is
unbounded in ${\cal O}(g^6)$.

\bm\section{$W$ Propagator in the Resonance Region}\ubm

A very recent work has extended the analysis to $W$ and quark
propagators in the resonance region.\cite{PaSi98}

One finds that a new problem  emerges: in the treatment of the
photonic corrections, conventional mass renormalization generates, in
NLO, a series in powers of $M\Gamma/(s-M^2)$,
which does not converge in the resonance region! Furthermore, it
diverges term-by-term at $s=M^2$. This problem is generally present
whenever the unstable particle is coupled to massless quanta. Aside
from the $W$, an interesting example is the QCD correction to a quark
propagator when the weak interactions are switched on, so that the quark
becomes unstable. A solution of this serious 
problem is presented in the framework of the complex pole
formalism.\cite{PaSi98}

\begin{figure}[t]
\centering
\mbox{\epsfig{file=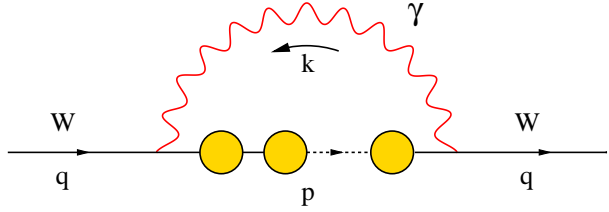,width=8cm,%
bbllx=0pt,bblly=0pt,bburx=509pt,bbury=175pt}}
\caption{A class of photonic corrections to the $W$ self-energy.
The inner solid and dashed lines and blobs represent transverse
$W$ propagators and self-energies.}
\label{f:one}
\end{figure}

In order to illustrate the difficulties emerging in the resonance
region when conventional mass renormalization is employed, we consider
the contribution of the transverse part of the $W$ propagator in the
loop of Fig.~\ref{f:one}, which contains $l$ self-energy insertions.

Calling
\be
        \Pi^{\scriptscriptstyle (T)}_{\mu \nu}(q) = t_{\mu \nu}(q) A(s),
\ee
the transverse $W$ self-energy, where $s\equiv q^2$ and
$t_{\mu \nu}(q)=$ $g_{\mu \nu} - q_\mu q_\nu/ q^2$,
the contribution $A_{\smallw \gamma}^{(l)}(s)$ from
Fig.~(1) to $A(s)$ is given
by
\bea
        A_{\smallw \gamma}^{(l)}(s) &=& i e^2(\mu)
        \,\frac{t_{\mu \nu}(q)}{(n-1)}\, \mu^{4-n}
                                        \nonumber \\
        \times & & \!\!\!\!\!\!\!\!\!\!\!\!
        \int \frac{d^n \!k}{(2\pi)^n}
                {\cal{D}}
                ^{{\scriptscriptstyle (} \gamma {\scriptscriptstyle )}}
                _{\rho \beta} (k)
        {\cal{D}}^{\scriptscriptstyle (W,T)}_{\lambda \alpha}(p)
        {\cal{V}}^{\rho \lambda \nu} {\cal{V}}^{\beta \alpha \mu}
        \left[\frac{A^{{\scriptscriptstyle (} s {\scriptscriptstyle )}}(p^2)}
                {p^2 -M^2 +i \epsilon} \right]^l ,
\label{eq:A-l-insertions}
\eea
where $p=q+k$ is the $W$ loop-momentum,
\be
        {\cal D}^{{\scriptscriptstyle (} \gamma {\scriptscriptstyle)}}
                _{\rho \beta} (k) = -\frac{i}{k^2}
        \left(g_{\rho \beta} + (\xig -1)
        \frac{k_\rho k_\beta}{k^2} \right),
\ee
\be
        {\cal D}^{\scriptscriptstyle (W,T)}_{\lambda \alpha}(p)=
        \frac{-i}{p^2 -M^2 +i\epsilon}
        \left(g_{\alpha \lambda} - \frac{p_\alpha p_\lambda}{p^2} \right),
\label{eq:Wprop-landau}
\ee
~
\be
        {\cal V}^{\beta \alpha \mu} = (2p-k)^\beta g^{\alpha \mu} +
        (2k-p)^\alpha g^{\beta \mu} -(k+p)^\mu g^{\beta \alpha},
\label{eq:vertex}
\ee
$\xig$ is the photon gauge parameter and
$A^{{\scriptscriptstyle (} s {\scriptscriptstyle )}}(p^2)$ stands for
the $W$ transverse self-energy with the conventional mass renormalization
subtraction:
\bea
        A^{{\scriptscriptstyle (} s {\scriptscriptstyle )}}(p^2) &=&
                        \mbox{Re}\left(A(p^2)-A(M^2)
                        \right) + i \mbox{Im}A(p^2)  \nonumber  \\
                &=& A(p^2) - A(M^2) + i\mbox{Im}A(M^2).
\eea
We note that each insertion of 
$A^{{\scriptscriptstyle (} s {\scriptscriptstyle )}}(p^2)$
is accompanied by an additional denominator
$[p^2-M^2 +i\epsilon]$. Thus, Eq.~(\ref{eq:A-l-insertions}) may be
regarded as the $l$th term in an expansion in powers of
$$
        \left[A(p^2) - A(M^2) + i\mbox{Im}A(M^2) \right]
        \left(p^2-M^2 +i\epsilon\right)^{-1}.
$$
As $A(p^2) - A(M^2) = O[g^2(p^2-M^2)]$ for $p^2 \approx M^2$, the 
contribution 
$[A(p^2) - A(M^2)]$ $(p^2-M^2 +i\epsilon)^{-1}$ is of $O(g^2)$ 
throughout the region of integration. However, as 
$i\mbox{Im}A(M^2) \approx -iM\Gamma$ is not subtracted, the combination
$i\mbox{Im}A(M^2)/(p^2-M^2 +i\epsilon)$ may lead to terms of $O(1)$
if the domain of integration $|p^2-M^2|$ \lsim $M \Gamma$ is important.
In fact, the contribution of $[i\mbox{Im}A(M^2)/(p^2-M^2 +i\epsilon)]^l$
to Eq.~(\ref{eq:A-l-insertions}) is, to leading order,
\be
        A_{\smallw \gamma}^{(l)}(s) =\frac{(-iM\Gamma)^l}{l!}
        \frac{d^l}{d(M^2)^l} A_{\smallw \gamma}^{(0)}(s) + \ldots,
\label{eq:A-l-insertions-Im}
\ee
where $A_{\smallw \gamma}^{(0)}(s)$ represents the diagram with no 
self-energy insertions and the dots indicate additional contributions
not relevant to our argument.

In the resonance region $|s-M^2|$ \lsim $M \Gamma$ the 
zeroth order propagator is inversely 
proportional to $(s-M^2+iM\Gamma) =O(g^2)$.
In NLO, contributions of $O[\alpha(s-M^2),\alpha M\Gamma]$ are therefore
retained but those of $O[\alpha(s-M^2)^2]$ are neglected.
Explicit evaluation of 
$A_{\smallw \gamma}^{(0)}(s)$ in NLO leads to 
\be
        A_{\smallw \gamma}^{(0)}(s) = \frac{\alpha}{2\pi}
        \left[(\xig-3)(s-M^2)\LM + \ldots\right].
\label{eq:A-0-insertions-partial}
\ee
Inserting Eq.~(\ref{eq:A-0-insertions-partial}) into 
Eq.~(\ref{eq:A-l-insertions-Im}) we obtain 
\bea
	A_{\smallw \gamma}^{(1)}(s) &=& \frac{\alpha}{2\pi}(\xig -3)
		\left( iM \Gamma \right) \left[ \LM + \fr{s}{M^2}
		\right] + \ldots,
			          \nonumber \\	
        A_{\smallw \gamma}^{(l)}(s) &=&\frac{\alpha}{2\pi}(\xig -3)
        \frac{(s-M^2)}{l(l-1)} 
        \left(\frac{-iM\Gamma\phantom{^2}}{s-M^2} \right)^{\!l} + \ldots, 
        \quad   (l \geq 2).
\label{eq:A-l-insertions-Im-again}
\eea
As in the resonance region all these terms contribute in NLO,
conventional mass renormalization leads in NLO to a series in powers
of $M\Gamma/(s-M^2)$, which does not converge in the resonance region.
Rather than generating contributions of higher order in $g^2$,
each successive self-energy insertion gives rise to a factor  
$-iM\Gamma/(s-M^2)$, which is nominally of $O(1)$ in the resonance region 
and furthermore diverges at $s=M^2$!
We note also that the use of Eq.~(\ref{eq:A-l-insertions-Im-again}) would lead
to power-behaved infrared divergences in $\re A(M^2)$ (mass counterterm) for
$l=2,4,6,\ldots$, and in the width for $l=3,5,7,\ldots$.

One possibility would be to resum the series 
$\sum^{\infty}_{l=0}A_{\smallw \gamma}^{(l)}(s)$ with 
$ A_{\smallw \gamma}^{(l)}(s)$ given by Eq.~(\ref{eq:A-l-insertions-Im}).
This would lead to 
\be
        \sum^{\infty}_{l=0} A_{\smallw \gamma}^{(l)}(s, M^2) = 
        A_{\smallw \gamma}^{(0)}(s, M^2-iM\Gamma) + \ldots,
\ee
or
\be
 	\sum^{\infty}_{l=0} A_{\smallw \gamma}^{(l)}(s)=\frac{\alpha}{2\pi}
        \left[(\xig-3)(s-M^2+iM\Gamma)\ln
        \left(\frac{M^2-iM\Gamma -s}{M^2-iM\Gamma}\right) + \ldots\right].
\label{eq:A-0-insertions-partial-again}
\ee
Even if one accepts these resummations rather than the usual term by
term expansions, the theoretical situation in the conventional
formalism is very unsatisfactory. 
In fact, in the conventional formalism, the $W$ propagator is inversely 
proportional to 
\be
        {\cal D}^{-1}(s) = s-M^2 +i M\Gamma 
		-\left( A(s)-A(M^2) \right) 
		-i M\Gamma \,\mbox{Re} A^\prime(M^2),
\label{eq:inverse_prop}
\ee
where $\Gamma$ is the radiatively corrected width and we have employed
its conventional expression
\be
	M \Gamma = -\mbox{Im}A(M^2)/[1-\mbox{Re}A'(M^2)].
\label{eq:usualwidth}
\ee 
The contribution of the 
$(s-M^2+iM\Gamma)\ln[(M^2-iM\Gamma -s)/(M^2-iM\Gamma)]$ term to 
${\cal D}^{-1}(s)$ is 
$$
        -\frac{\alpha}{2\pi} (\xig-3) \left[(s-M^2+iM\Gamma)\ln
        \left(\frac{M^2-iM\Gamma -s}{M^2-iM\Gamma}\right)
        +iM\Gamma \left(1+i\frac{\pi}{2} \right) \right]
$$
and we note that the last term is a gauge-dependent contribution not 
proportional to the zeroth order term $s-M^2+iM\Gamma$. 
As a consequence, in NLO the pole position is shifted to 
${\widetilde{M}}^2 -i\widetilde{M} \widetilde{\Gamma}$, where
\bea
	{\widetilde{M}}^2 &=& M^2[1-(\alpha/4)(\xig-3)(\Gamma/M)],
\label{eq:Mtilde}				\\
	\widetilde{\Gamma} &=& \Gamma [1-(\alpha/2\pi)(\xig-3)].
\label{eq:Gammatilde}	
\eea
As the pole position is gauge-invariant, so must be $\widetilde{M}$ and
$\widetilde{\Gamma}$. Furthermore, in terms of $\widetilde{M}$ and
$\widetilde{\Gamma}$, ${\cal D}^{-1}(s)$ retains the Breit-Wigner
structure. Thus, in a resonance experiment $\widetilde{M}$ and
$\widetilde{\Gamma}$ would be identified with the mass and width of $W$.

The relation $\widetilde{\Gamma} = \Gamma [1-(\alpha/2\pi)(\xig-3)]$
leads to a contradiction: the measured, gauge-independent, width 
$\widetilde{\Gamma}$ would differ from the theoretical value
$\Gamma$ by a gauge-dependent quantity in NLO! This contradicts the
premise of the conventional formalism that $\Gamma$, defined in
Eq.~(\ref{eq:usualwidth}), is the radiatively corrected width and is,
furthermore, gauge-independent. We can anticipate that the root of
this clash between the resummed expression and the conventional
definition of width is that the latter is only an approximation.
In particular, it is not sufficiently accurate when non-analytic
contributions are considered.

A good and consistent formalism may circumvent awkward resummations 
of non-convergent series and should certainly avoid the above
discussed contradictions.  
To achieve this, we return to the transverse dressed $W$ propagator,
inversely proportional to $p^2-M_0^2 -A(p^2)$. In the conventional
mass renormalization one eliminates $M^2_0$ by means of the expression
$M^2_0=M^2 -\mbox{Re}A(M^2)$ (Cf. Eq.~(\ref{eq:oms})). 
An alternative possibility is to
eliminate $M^2_0$ by means of $M^2_0=\sov -A(\sov)$
(Cf. Eq.~(\ref{eq:pol})).
The dressed propagator in the loop integral is inversely proportional
to $p^2 -\sov -[A(p^2)-A(\sov)]$. Its expansion about $p^2 -\sov$
generates in Fig.~(1) a series in powers of 
$[A(p^2)-A(\sov)]/(p^2-\sov)$.
As $A(p^2)-A(\sov)= O[g^2(p^2-\sov)]$ when the loop momentum is in the
resonance region, $[A(p^2)-A(\sov)]/(p^2 -\sov)$ is $O(g^2)$ 
throughout the domain of
integration. Thus, each successive self-energy insertion leads now to
terms of higher order in $g^2$ without awkward non-convergent
contributions.
In this modified strategy, the zeroth order propagator
in Eq.~(\ref{eq:Wprop-landau}) is replaced by 
\be
        {\cal D}^{\scriptscriptstyle (W,T)}_{\alpha \lambda }(p)=
        \frac{-i}{p^2 -\sov}
        \left(g_{\alpha \lambda} - \frac{p_\alpha p_\lambda}{p^2} \right).
\label{eq:Wprop-landau-sov}
\ee
The poles in the $k^0$ complex plane remain in the same quadrants as
in Feynman's prescription and Feynman's contour integration or 
Wick's rotation can be carried out.
$A_{\smallw \gamma}^{(0)}(s)$, Fig.~(1) without loop insertions, 
now leads directly to 
\be
        A_{\smallw \gamma}^{(0)}(s) = \frac{\alpha}{2\pi}
        \left[(\xig-3)(s-\sov)\LS + \ldots\right],	
\label{eq:A-0-insertions-partial-new}
\ee
which has the same structure as the expression we obtained in the
conventional approach after resumming a non-convergent series.
$A_{\smallw \gamma}^{(l)}(s)$ ($l\geq 1$), the contributions 
with $l$ insertions in Fig.~(1), are now of $O(\alpha g^{2l})$, 
the normal situation in perturbative expansions.
The $W$ propagator in the modified formalism is inversely
proportional to $s-\sov- [A(s)-A(\sov)]$. As $A_{\smallw
\gamma}^{(0)}(s)$ is now proportional to $s-\sov$, the pole position
is not displaced, the gauge-dependent contributions factorize as
desired, and the above discussed pitfalls are avoided. 
$A_{\smallw \gamma}^{(l)}(s)$ leads now to
contributions to $[A(s)-A(\sov)]$ of order  
$O[(s-\sov)\alpha g^{2l}] = O[\alpha g^{2(l+1)}]$ in the resonance
region and can therefore be neglected in NLO for $l\geq 1$.
We note that the $\ln [(\sov-s)/\sov]$ term in 
Eq.~(\ref{eq:A-0-insertions-partial-new}) cancels for $\xig=3$, the gauge
introduced by Fried and Yennie in Lamb-shift calculations.\cite{Fried-Yennie} 

The remaining contributions to $A(s)$ from the photonic diagrams,
including those from the longitudinal part of the $W$ propagator in 
Fig.~(1), and from the diagrams involving the unphysical scalars
$\phi$ and the ghost $C_\gamma$, have no singularities at $s=M^2$
and can therefore be studied with conventional methods. 

Calling $A^{\gamma}(s)$ the overall contribution of the 
one-loop photonic diagrams to the transverse $W$ self-energy, 
in the modified
formulation the relevant quantity in the correction to the $W$
propagator is $A^{\gamma}(s) - A^{\gamma}(\sov)$. 
In general $R_\xi$ gauge, we find in NLO
\bea 
        \lefteqn{\Agamma(s) - \Agamma(\sov) = 
                \frac{\alpha(m_2)}{2\pi} (s-\sov)
                \left\{\delta 
                \left(\frac{\xiw}{2}-\frac{23}{6}\right)
                +\frac{34}{9} -2\LS
                                \right.}                \nonumber \\
        & & \!\!\!\!\!\!\!\!\!\!
        -\left(\xiw-1\right)\left[\fr{\xiw}{12} -
                \left(1-\fr{(\xiw-1)^2}{12}\right)
                \ln \!\left(\fr{\xiw-1}{\xiw}\right) \right]      
                -\left(\fr{11}{12} 
                -\fr{\xiw}{4} \right) \ln\xiw
							\nonumber \\
        & & \!\!\!\!\!\!\!\!\!\!
        \left. +\left(\xig-1\right)\left[
                \fr{\delta}{2}+\fr{1}{2}
                +\LS+\fr{(\xiw^2-1)}{4}\ln\!\left(\fr{\xiw-1}{\xiw}\right) 
                -\fr{\ln\xiw}{4}+\fr{\xiw}{4} \right] \right\},
							\nonumber \\
\label{eq:Agamma-Agamma}                                                       
\eea
where $\delta = (n-4)^{-1} + (\gamma_E -\ln 4\pi)/2$
and we have set $\mu=m_2$.
Of particular interest in Eq.~(\ref{eq:Agamma-Agamma})
is the log term 
$$
	\frac{\alpha(m_2)}{2\pi} \left(\xig-3\right)
	\left(s-\sov\right)\LS,
$$
which is independent of $\xiw$ but is proportional to $(\xig-3)$.
The logarithm $\ln(\xiw-1)$ in Eq.~(\ref{eq:Agamma-Agamma})
contains an imaginary contribution 
$-i\pi \theta(1-\xiw)$. This can be understood from the observation
that, for $\xiw<1$, a $W$ boson of mass $s=M^2$ has non-vanishing
phase space to ``decay'' into a photon and particles of mass $M^2\xiw$.

\section{Gauge Dependence of the On-Shell Mass}

The difference between the pole mass $m_1$, defined in
Eq.~(\ref{eq:lep}), and the conventional on-shell mass $M$, 
defined in  Eq.~(\ref{eq:oms}), is
\be
        M^2 -m_1^2 = \mbox{Re} A(M^2) - \mbox{Re} A(\sov) -\Gamma_2^2.
\label{eq:deltaM}
\ee
The contribution of the $(s-\sov)\ln[(\sov-s)/\sov]$ term to the r.h.s. 
of Eq.~(\ref{eq:deltaM}) is
\bea
	& &\frac{\alpha(m_2)}{2\pi}\left(\xig -3\right)
	   \left[\left(M^2-m_2^2\right) 
		\mbox{Re}\ln \!\left(\frac{\sov-M^2}{\sov}\right)
		-m_2 \Gamma_2 \mbox{Im}\ln \! 
		\left(\frac{\sov-M^2}{\sov}\right)
	   \right]				\nonumber \\
	& \approx & \frac{\alpha(m_2)}{2\pi}\left(\xig -3 \right)
	   \left[\left(M^2-m_1^2\right)
		\mbox{Re}\ln \! \left(\frac{\sov-M^2}{\sov}\right)
		+m_2 \Gamma_2 \frac{\pi}{2}\right].
\eea
In $\mbox{Im}\ln[(\sov-M^2)/\sov]$ we have approximated 
$M^2 \approx m_1^2$ and used the fact that $\theta=-\pi/2$ for
$s= m_1^2$.
Thus, we have 
\be
        M^2 -m_1^2 = \fr{\alpha(m_2)}{4}(\xig-3)m_2\Gamma_2+ \ldots,
\label{eq:deltaMagain}
\ee
where the dots indicate additional contributions. We note that this last
equation corresponds to our previous result from the propagator, 
Eq.~(\ref{eq:Mtilde}), with the identification 
$\widetilde{M} \rightarrow m_1$.
In particular, Eq.~(\ref{eq:deltaMagain}) leads to 
$m_1-M=\alpha(m_2) \Gamma_2/4 \approx 4 \mev$ in the frequently 
employed 't Hooft-Feynman gauge $(\xi_i=1)$, and to $\approx 6\mev$
in the Landau gauge $(\xi_i=0)$.
The contribution to $M^2-m_1^2$ from the term proportional to 
$(s-\sov)(\xig-1)(\xiw^2-1)\ln(\xiw-1)$ (Cf.~Eq.~(\ref{eq:Agamma-Agamma})) 
is $(\alpha/8)(\xig-1)M\Gamma(\xiw^2-1)\,\theta(1-\xiw)$, which is 
unbounded in $\xig$ but restricted to $\xiw<1$.
In analogy with the $Z$ case, 
there are also bounded gauge-dependent contributions to $m_1-M$
arising from non-photonic diagrams in the restricted range
$\mw>\mz\sqrt{\xiz}+\mw\sqrt{\xiw}$ or
$\sqrt \xiz \leq \cos \theta_{\mbox{\footnotesize{w}}}[1-\sqrt \xiw]$,
and from the photonic corrections proportional to 
$(\xiw-1)\ln[(\xiw-1)/\xiw]$ (Cf. Eq.~(\ref{eq:Agamma-Agamma})).

The following observation is appropriate at this point. In calculating
$\Delta r\ $\cite{Si80} (and its $\overline{\rm MS}$ counterparts,
$\Delta\hat r\ $\cite{Si89-DFS91} and $\Delta\hat r_W\ $\cite{FaSi90}),
one must consider the
$W$ mass counterterm. In the complex pole formalism, the mass
counterterm is $\re A(\bar s)$ and we see that the contribution from
Eq.~(\ref{eq:A-0-insertions-partial-new}) vanishes exactly.
If one employs instead the conventional mass counterterm
$\re A(M^2)$, the resummed expression of 
Eq.~(\ref{eq:A-0-insertions-partial-again}) gives an unbounded
gauge-dependent contribution $(\alpha/4)(\xi_\gamma-3)M\Gamma$.
The same result is obtained if one restricts one-self to
$A_{\smallw\gamma}^{(0)}+A_{\smallw\gamma}^{(1)}$
(Cf.\ Eqs.~(\ref{eq:A-0-insertions-partial}) and
(\ref{eq:A-l-insertions-Im-again})), rather
than Eq.~(\ref{eq:A-0-insertions-partial-again}), and evaluates the imaginary
part of the logarithm at $s=M^2$
using the $i\epsilon$ prescription. One should eliminate these gauge dependent
terms by means of the replacement
$M^2-(\alpha/4)(\xi_\gamma-3)M\Gamma=m_1^2$
and identify $m_1$ with the measured
mass. On the other hand, if we again retain only
$A_{\smallw\gamma}^{(0)}+A_{\smallw\gamma}^{(1)}$ but regulate the
logarithm with an infinitesimal photon mass $\lambda$ when $s=M^2$,
$A_{\smallw\gamma}^{(1)}(M^2)$ is purely
imaginary, so that it does not contribute to $\re A(M^2)$, and the
above-mentioned gauge dependence in $\re A(M^2)$ and
Eq.~(\ref{eq:deltaMagain}) does not arise.

\bm\section{Overall Corrections to $W$ Propagator in the Resonance Region}\ubm

In contrast with the photonic corrections, the non-photonic
contributions $A_{np}(s)$ to $A(s)$ are analytic around $s=\sov$.   
In NLO we can therefore write 
\be
	A_{np}(s)-A_{np}(\sov) = (s-\sov)A_{np}'(m_2^2)+ \ldots,
\ee
where the dots indicate higher-order contributions. 

In the resonance region, and in NLO, the transverse $W$ propagator 
becomes
\be
	{\cal D}^{\scriptscriptstyle (W,T)}_{\alpha \beta}(q)=
        \frac{-i \left(g_{\alpha \beta} - q_\alpha q_\beta/ q^2\right)}
	     {\left(s-\sov\right)
		\left[1-A_{np}'(m_2^2) 
		-\frac{\alpha(m_2)}{2\pi}F(s,\sov,\xig,\xiw) \right]},
\label{eq:fullWprop-landau}
\ee
where $s=q^2$ and $F(s,\sov,\xig,\xiw)$ is the expression between
curly brackets in Eq.~(\ref{eq:Agamma-Agamma}). An alternative
expression, involving an $s-$dependent width, can be obtained by
splitting $A_{np}'$ into real and imaginary parts, and the latter into 
fermionic Im$A'_f$ and bosonic Im$A'_b$ contributions. 
Neglecting very small scaling violations, we have 
\be
	\mbox{Im} A'_f(m_2^2) \approx \mbox{Im} A_f(m_2^2)/m_2^2
			\approx -\Gamma_2/m_2
\ee
and 
\be
	{\cal D}^{\scriptscriptstyle (W,T)}_{\alpha \beta}(q)=
        \frac{-i \left(g_{\alpha \beta} - q_\alpha q_\beta/ q^2\right)}
	     	{\left(s-m_1^2 +is\frac{\Gamma_1}{m_1}\right)
	     	\left[1-\mbox{Re}A_{np}'(m_1^2)-i\mbox{Im}A_b'(m_1^2)
		-\frac{\alpha(m_1)}{2\pi}F \right] },
\label{eq:fullWprop-landau-again}
\ee
where $\Gamma_1/m_1=\Gamma_2/m_2$. $\mbox{Im}A_b'(m_1^2)$ is non-zero
and gauge-dependent in the subclass of gauges that satisfy 
$\sqrt \xiz \leq \cos \theta_{\mbox{\footnotesize{w}}}[1-\sqrt \xiw]$.
Otherwise $\mbox{Im}A_b'(m_1^2)$
vanishes. Although $m_1$ and $\Gamma_1$ are gauge-invariant, 
$\mbox{Re}A'_{np}(m_1^2)$, $\mbox{Im}A'_{np}(m_1^2)$
 and $F$ are gauge-dependent. In physical
amplitudes, such gauge-dependent terms cancel against contributions
from vertex and box diagrams.
The crucial point is that the gauge-dependent contributions in 
Eq.~(\ref{eq:fullWprop-landau-again}) factorize so that such
cancellations can take place and the position of the complex pole is
not displaced.
 
\section{QCD Corrections to Quark Propagators in the Resonance Region}

In pure QCD quarks are stable particles, but  they become unstable
when weak interactions are switched on. As we anticipate similar 
problems to those in the $W$ case, we work from the outset in the
complex pole formulation.
Calling $\mov = m -i \Gamma/2$ the position of the complex pole,
$\Gamma$ arises from the weak interactions. If we treat $\Gamma$ to
lowest order (LO), but otherwise neglect the remaining weak interactions
contributions to the self-energy, the dressed quark propagator can be
written
\be
	S'_F(\qsl) = \frac{i}{\qsl -\mov - 
		\left( \Sigma(\qsl) - \Sigma(\mov) \right)},  
\ee
where $\Sigma(\qsl)$ is the pure QCD contribution.
In NLO, in the resonance region, one finds
\be
	S'_F(\qsl) = \frac{i}{\left(\qsl -\mov \right)} 
		\left\{1-
		\frac{\alpha_s(m)}{3\pi} \left[2\left(\xi_g-3\right) 
			\ln \! \left(\frac{\mov^2-q^2}{\mov^2}\right)+
			2 \delta \xi_g \right]+\ldots \right\}^{-1},
\label{eq:S'_F}
\ee
where $\xi_g$ is the gluon gauge parameter and we have set $\mu=m$.
As in the $W-$propagator case, we see that the logarithm vanishes in
the Fried-Yennie gauge $\xi_g =3$.

\section{Conclusions for First Subject}

The conclusions can be summarized in the following points.
i) Conventional mass renormalization, when applied to photonic and
gluonic diagrams, leads to a series in powers of $M\Gamma/(s-M^2)$
in NLO which does not converge in the resonance region.
ii) In principle, this problem can be circumvented by a resummation
procedure.
iii) Unfortunately, the resummed expression leads to an inconsistent
answer, when combined with the conventional definition of width.
This is not too surprising, as the traditional expression of width 
treats the unstable particle as an asymptotic state, which is clearly
only an approximation.
iv) An alternative treatment of the resonant propagator is
discussed, based on the complex-valued pole position 
$\sov= M_0^2+A(\sov)$.
The non-convergent series in the resonance region and the potential
infrared divergences in $\Gamma$ and $M$ are avoided by employing 
$(p^2-\sov)^{-1}$ rather than $(p^2-M^2)^{-1}$ in the Feynman integrals. 
The one-loop diagram leads now directly to the resummed expression of
the conventional approach, while the multi-loop expansion generates
terms which are genuinely of higher order. The non-analytic terms and
the gauge-dependent corrections cause no problem because they are
proportional to $s-\sov$ and therefore exactly factorize. 
v) The presence of $\sov$ in $\ln[(\sov-s)/\sov]$ removes the problem
of apparent on-shell singularities.
vi) The gauge dependence of the on-shell definition of mass for $W$ bosons
present new features discussed in Section~3.

\section{Differences Between the Pole and On-Shell Masses and Widths of the
Higgs Boson}

Eqs.~(\ref{eq:oms}--\ref{eq:mgp}) are also applicable to unphysical scalars.
In this case, $A(s)$ is the corresponding self-energy.
As explained in the Introduction,
if one expands Eq.~(\ref{eq:mgp}) in powers of $m_2\Gamma_2$ about $m_2^2$,
the result agrees with Eq.~(\ref{eq:oms}) in NLO.
The leading differences, which occur in next-to-next-to-leading order (NNLO),
have been studied.\cite{prl}
One finds:
\begin{eqnarray}
\label{eq:lin}
\frac{M-m_2}{m_2}&=&-\frac{\Gamma_2}{2m_2}\im A^\prime(m_2^2)+{\cal O}(g^6),
\\
\frac{\Gamma-\Gamma_2}{\Gamma_2}&=&
\im A^\prime(m_2^2)\left(\frac{\Gamma_2}{2m_2}+\im A^\prime(m_2^2)\right)
-\frac{m_2\Gamma_2}{2}\im A^{\prime\prime}(m_2^2)+{\cal O}(g^6),
\nonumber
\end{eqnarray}
where $g^2$ is a generic coupling of ${\cal O}(\Gamma_2/m_2)$.
As the right-hand sides of Eq.~(\ref{eq:lin}) are of ${\cal O}(g^4)$, we may 
evaluate them using the LO expressions for $\Gamma_2$,
$\im A^\prime(m_2^2)$, and $\im A^{\prime\prime}(m_2^2)$.

In the Higgs-boson case, the one-loop bosonic contribution to $\im A(s)$ in
the $R_\xi$ gauge is given by
\begin{eqnarray}
\label{eq:bos}
\im A_{\rm bos}(s)&=&\frac{G}{4}s^2\left[-\left(1-\frac{4M_W^2}{s}
+\frac{12M_W^4}{s^2}\right)\left(1-\frac{4M_W^2}{s}\right)^{1/2}
\theta(s-4M_W^2)
\right.
\\
&&{}+\left.
\left(1-\frac{M_H^4}{s^2}\right)\left(1-\frac{4\xi_WM_W^2}{s}\right)^{1/2}
\theta(s-4\xi_WM_W^2)
+\frac{1}{2}(W\to Z)\right],
\nonumber
\end{eqnarray}
where $G=G_\mu/(2\pi\sqrt2)$, $\xi_W$ is a gauge parameter, $(W\to Z)$
represents the sum of the preceding terms with the substitutions $M_W\to M_Z$
and $\xi_W\to\xi_Z$, and we have omitted gauge-invariant terms proportional to
$\theta(s-4M_H^2)$.
The one-loop contribution due to a fermion $f$ is
\begin{equation}
\im A_f(s)=-\frac{G}{2}sN_fm_f^2\left(1-\frac{4m_f^2}{s}\right)^{3/2}
\theta(s-4m_f^2),
\label{eq:fer}
\end{equation}
where $N_f=1$ (3) for leptons (quarks).
As expected, Eq.~(\ref{eq:bos}) is gauge invariant if $s=M_H^2$, but it
depends on $\xi_W$ and $\xi_Z$ off-shell.
The $\xi_W$ dependence in Eq.~(\ref{eq:bos}) is due to the fact that a Higgs 
boson of mass $s^{1/2}>2\xi_W^{1/2}M_W$ has non-vanishing phase space to 
``decay" into a pair of ``particles" of mass $\xi_W^{1/2}M_W$.
The first term in Eq.~(\ref{eq:bos}) can be verified by a very simple 
argument:{}\cite{rin} in Eq.~(\ref{eq:bos}),
only the unphysical scalar excitations have $M_H$-dependent couplings with the
Higgs boson; therefore, if the unphysical particles decouple, which happens
for $\xi_W>s/(4M_W^2)$ and similarly for the $Z$ boson, $\im A(s)$ can be
obtained by substituting $M_H^2\to s$ in the well-known expressions for the
Higgs-boson partial widths multiplied by $M_H$.
Using Eqs.~(\ref{eq:bos}) and (\ref{eq:fer}), we find
$\im A_{\rm bos}^\prime(m_2^2)$, $\im A_{\rm bos}^{\prime\prime}(m_2^2)$,
$\im A_{\rm fer}^\prime(m_2^2)$, and $\im A_{\rm fer}^{\prime\prime}(m_2^2)$.
This permits us to 
evaluate Eq.~(\ref{eq:lin}).
We also wish to evaluate $(M^{\rm PT}-m_2)/m_2$ and
$(\Gamma^{\rm PT}-\Gamma_2)/\Gamma_2$, where $M^{\rm PT}$ and
$\Gamma^{\rm PT}$ are the pinch-technique (PT) on-shell mass and width 
obtained from Eq.~(\ref{eq:oms}) by employing the PT self-energy $a(s)$.
We recall that the PT is a prescription that combines conventional 
self-energies with ``pinch parts" from vertex and box diagrams in such a 
manner that the modified self-energies are independent of $\xi_i$ 
($i=W,Z,\gamma$) and exhibit desirable theoretical properties.\cite{cor}
In the Higgs-boson case, $\im a(s)$ can be extracted from the
literature.\cite{pil}

\begin{figure}[t]
\centering
\mbox{\epsfig{file=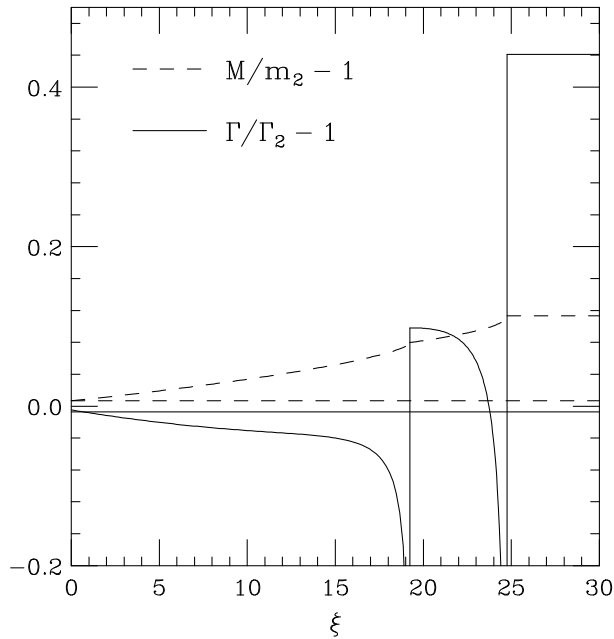,width=8cm,%
bbllx=123pt,bblly=168pt,bburx=539pt,bbury=604pt}}
\caption{Relative deviations of $M$ and $\Gamma$ from $m_2$ and $\Gamma_2$,
respectively, as functions of $\xi=\xi_W=\xi_Z$ in the $R_\xi$ gauge, assuming
$m_2=800$~GeV.
The horizontal lines across the figures indicate the corresponding deviations
in the PT framework.
The two abysses are associated with the unphysical thresholds
$\xi=m_2^2/(4M_Z^2),m_2^2/(4M_W^2)$, where the expansions in
Eq.~(\ref{eq:lin}) obviously fail.}
\label{f:two}
\end{figure}

Identifying $M_H$ with $m_2$ and, for simplicity, setting $\xi=\xi_W=\xi_Z$, 
we have evaluated
$(M-m_2)/m_2$ and $(\Gamma-\Gamma_2)/\Gamma_2$
as functions of $\xi$, for three values of $m_2$.
We have employed $M_W=80.375$~GeV, $M_Z=91.1867$~GeV, and $m_t=175.6$~GeV, and
have neglected contributions from fermions other than the top quark.
For small Higgs mass ($m_2=200$~GeV), we find that,
aside from the neighborhoods of unphysical thresholds where the expansion of
Eq.~(\ref{eq:lin}) fails, $M$ and $\Gamma$ remain 
numerically very close to $m_2$ and $\Gamma_2$.
In the intermediate case ($m_2=400$~GeV), the relative differences reach
0.6\% in the mass and 3.3\% in the width.
However, for a heavy Higgs boson ($m_2=800$~GeV), the differences become very
large, reaching 11\% in the mass and 44\% in the width
(see Fig.~\ref{f:two}).
The largest differences occur for $\xi>m_2^2/(4M_W^2)$, i.e., when the
unphysical excitations decouple, a range that includes the unitary gauge.
We recall that the latter retains only the physical degrees of freedom and, in 
this sense, it may be regarded as the most physical of all gauges.
The large effects can be easily understood from Eq.~(\ref{eq:bos}).
If $\xi>s/(4M_W^2)$, the second term in Eq.~(\ref{eq:bos}) does not
contribute, so that $\im A_{\rm bos}(s)\propto s^2$.
For a heavy Higgs boson, this implies large values of $\im A^\prime(m_2^2)$
and $\im A^{\prime\prime}(m_2^2)$.
For $\xi<s/(4M_Z^2)$, the gauge-dependent terms contribute and cancel the 
leading $s^2$ dependence of $\im A_{\rm bos}(s)$, so that the magnitudes of
$\im A^\prime(m_2^2)$ and $\im A^{\prime\prime}(m_2^2)$ drop sharply and the 
differences become much smaller.
Of course, the 44\% effect in the width for $\xi>m_2^2/(4M_W^2)$ may cast 
doubts on the convergence of the expansions in Eq.~(\ref{eq:lin}).
We interpret this finding as an indication of large corrections rather than a 
precise evaluation of $(\Gamma-\Gamma_2)/\Gamma_2$.

Our results go beyond those reported in the literature.\cite{wil,swi,ghi,agh}
The reason is easy to understand:
in these papers,\cite{wil,swi,ghi,agh} the limits $M_W\to0$ and $g\to0$ are
simultaneously considered keeping the Higgs self-coupling
$\lambda\propto g^2M_H^2/M_W^2$ fixed.
If the gauge parameter $\xi$ is also kept fixed, the gauge dependence of 
Eq.~(\ref{eq:bos}) is lost, and one obtains an $s$-independent result for
$\im A_{\rm bos}(s)$, which does not contribute to the right-hand sides of
Eq.~(\ref{eq:lin}).
Thus, the above approximation, although interesting and useful, does not
exhibit the gauge dependence and the large effects discussed here.

From the horizontal lines across Fig.~\ref{f:two}, we see that the 
PT mass and width remain very close to $m_2$ and $\Gamma_2$,
the maximum departures being 0.7\% for $M^{\rm PT}$ and $-0.7\%$ for
$\Gamma^{\rm PT}$.
The differences vary somewhat if $M$ and $\Gamma$ are compared with $m_1$ and
$\Gamma_1$.
Through ${\cal O}(g^6)$, $(M-m_1)/m_1$ and $(\Gamma-\Gamma_1)/\Gamma_1$ are 
obtained from $(M-m_2)/m_2$ and $(\Gamma-\Gamma_2)/\Gamma_2$ by subtracting
the gauge-invariant term $\Gamma_2^2/(2m_2^2)$.
For $m_2=800$~GeV, $(M-m_1)/m_1$ and $(\Gamma-\Gamma_1)/\Gamma_1$ amount to
5.6\% and 38\% in the unitary gauge (rather than 11\% and 44\%) and to
$-4.8\%$ and $-6.6\%$ in the 't~Hooft-Feynman gauge (rather than 0.9\% and 
$-0.8\%$).
For the same value of $m_2$, the differences $(M^{\rm PT}-m_1)/m_1$ and
$(\Gamma^{\rm PT}-\Gamma_1)/\Gamma_1$ are $-5.1\%$ and $-6.5\%$ (rather than
0.7\% and $-0.7\%$).

\section{Mass and Width of a Heavy Higgs Boson}

As the gauge-dependent effects in the width and the mass occur in NNLO,
one would like to examine what happens when the widths, rather than
differences, are evaluated to this order.\cite{plb} Such calculations are, in
fact, available in the recent literature in the heavy-Higgs approximation
(HHA): $g\to0$, $M_W\to0$, with the Higgs quartic coupling
$\lambda\propto g^2M_H^2/M_W^2$ fixed. What happens to the
gauge dependence in this limit?

For finite values of the gauge parameters, $\xi_W$ and $\xi_Z$,
$\xi_WM_W^2,\xi_ZM_Z^2\to0$ as $M_W^2,M_Z^2\to0$.
Therefore, the second term in Eq.~(\ref{eq:bos}) contributes and cancels the
leading $s$ dependence of the first one.
Thus, for finite values of $\xi_W$ and $\xi_Z$, one obtains in the HHA
\begin{equation}
\im A(s)=-\frac{3}{8}GM^4\qquad
\mbox{($R_\xi$ gauge)},
\label{eq:rxi}
\end{equation}
independent of $s$.
Denoting by $M_\xi$ and $\Gamma_\xi$ the on-shell mass and width in the 
$R_\xi$ gauge (defined for finite values of $\xi_W$ and $\xi_Z$) and applying 
henceforth the HHA, Eqs.~(\ref{eq:lin}) and (\ref{eq:rxi}) lead to
\begin{equation}
\frac{M_\xi}{m_2}=1+{\cal O}(\lambda^3),\qquad
\frac{\Gamma_\xi}{\Gamma_2}=1+{\cal O}(\lambda^3).
\label{eq:mrxi}
\end{equation}
Instead, in the unitary gauge, one first takes the limit
$\xi_W,\xi_Z\to\infty$, in which case the term proportional to
$\theta(s-4\xi_WM_W^2)$ in Eq.~(\ref{eq:bos}) does not contribute, and one 
finds
\begin{equation}
\im A(s)=-\frac{3}{8}Gs^2\qquad
\mbox{(unitary gauge)}.
\label{eq:uni}
\end{equation}
Denoting by $M_u$ and $\Gamma_u$ the on-shell quantities in the unitary gauge,
Eqs.~(\ref{eq:lin}) and (\ref{eq:uni}) tell us that
\begin{equation}
\frac{M_u}{m_2}=1+\frac{9}{64}G^2m_2^4+{\cal O}(\lambda^3),\qquad
\frac{\Gamma_u}{\Gamma_2}=1+\frac{9}{16}G^2m_2^4+{\cal O}(\lambda^3).
\label{eq:muni}
\end{equation}
Comparison of Eq.~(\ref{eq:mrxi}) with Eq.~(\ref{eq:muni}) shows that, in the 
HHA, the leading gauge dependence of the on-shell mass or width reduces to a 
discontinuous function, with one value corresponding to finite
$\xi_W$ and $\xi_Z$, and the other one to the unitary gauge.
It should be pointed out, however, that for finite and large values of $\xi_W$
and $\xi_Z$, the limit $\xi_WM_W^2,\xi_ZM_Z^2\to0$ is not realistic within the 
SM, and must be regarded as a special feature of the HHA.

The relation between $\Gamma_3$ and $m_3$ was first obtained in NNLO by 
Ghinculov and Binoth,\cite{agh} with a numerical evaluation of the expansion
coefficients.
We have independently derived this expansion.
The relation between $m_3$ and $M_\xi$ was first given analytically in NNLO by 
Willenbrock and Valencia,\cite{swi} an expansion that we have also verified.
As the connection between the three pole parametrizations $(m_1,\Gamma_1)$,
$(m_2,\Gamma_2)$, and $(m_3,\Gamma_3)$ is known exactly from
Eqs.~(\ref{eq:pol})--(\ref{eq:p3}), and the relations of $(M_\xi,\Gamma_\xi)$
and $(M_u,\Gamma_u)$ with $(m_2,\Gamma_2)$ are given to the required accuracy 
in Eqs.~(\ref{eq:mrxi}) and (\ref{eq:muni}), we readily find in NNLO the 
expansions of $\Gamma_i$ ($i=1,2,3,\xi,u$) in terms of $m_i$ in the three pole
and two on-shell schemes discussed above to be
\begin{equation}
\Gamma_i=\frac{3}{8}Gm_i^3\left[1+a\frac{Gm_i^2}{\pi}
+b_i\left(\frac{Gm_i^2}{\pi}\right)^2\right],
\label{eq:gam}
\end{equation}
where
\begin{eqnarray}
a&=&\frac{5}{4}\zeta(2)-\frac{3}{4}\pi\sqrt3+\frac{19}{8},\qquad
b_2=b_\xi=0.96923(13),\nonumber\\
b_1&=&b_2-\frac{9\pi^2}{64},\qquad
b_3=b_2-\frac{9\pi^2}{128},\qquad
b_u=b_2+\frac{9\pi^2}{64}.
\end{eqnarray}
For the ease of notation, we have put $m_\xi=M_\xi$ and $m_u=M_u$.
Here, we have adopted the value for $b_\xi$ from Frink {\it et al}.\cite{fri}
It slightly differs from the value $0.97103(48)$.\cite{ghi,agh}
Although the difference is larger than the estimated errors, it amounts to 
less than 0.7\% in the coefficients $b_i$, which is unimportant for our
purposes.
On the other hand, $m_i$ ($i=1,3,\xi,u$) are related to $m_2$ by
\begin{equation}
m_i=m_2\left[1+c_i\left(\frac{Gm_2^2}{\pi}\right)^2
+d_i\left(\frac{Gm_2^2}{\pi}\right)^3\right],
\label{eq:mas}
\end{equation}
where
\begin{eqnarray}
c_1&=&\frac{9\pi^2}{128},\qquad
c_3=\frac{9\pi^2}{512},\qquad
c_\xi=0,\qquad
c_u=\frac{9\pi^2}{64},\nonumber\\
d_1&=&\frac{9\pi^2}{64}a,\qquad
d_3=\frac{9\pi^2}{256}a,\qquad
d_\xi=\frac{9\pi^2}{128}\left[-\frac{5}{4}\zeta(2)+\frac{\pi}{3}\sqrt3
+\frac{7}{8}\right],
\end{eqnarray}
while $d_u$ is currently unknown.
In the case of $i=3$, Eq.~(\ref{eq:gam}) agrees with Eq.~(10) of 
Ref.~17 up to the numerical difference in $b_3$ discussed above.

\begin{figure}[t]
\centering
\mbox{\epsfig{file=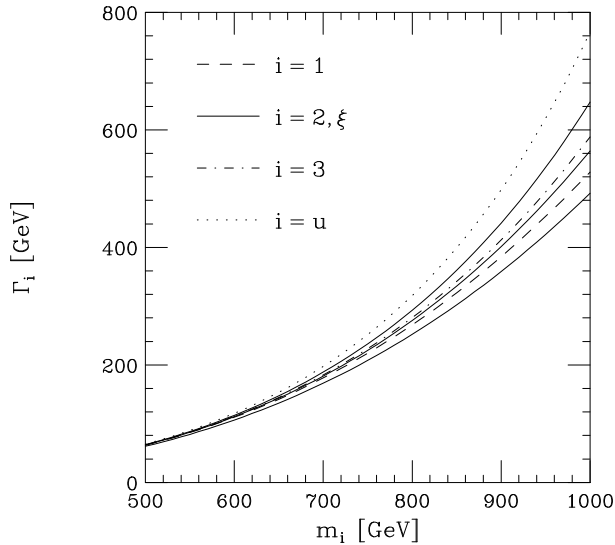,width=8cm,%
bbllx=54pt,bblly=168pt,bburx=548pt,bbury=608pt}}
\caption{Higgs-boson widths $\Gamma_i$ ($i=1,2,3,\xi,u$) as functions of the
corresponding masses $m_i$ in the various pole and on-shell schemes.
The down-most and middle solid lines correspond to the LO and NLO results,
which are common to all renormalization schemes, while the up-most one refers
to the NNLO result for $i=2,\xi$.}
\label{f:thr}
\end{figure}

We see from Eq.~(\ref{eq:gam}) that all the width expansions have the same 
LO and NLO coefficients.
This is due to the fact that the on-shell and pole widths only differ in
NNLO\cite{prl} and that the relations (\ref{eq:mas}) among the masses do not
involve terms linear in $Gm_2^2/\pi$.
It is also interesting to note that the on-shell mass $M_{\rm PT}$ and width
$\Gamma_{\rm PT}$, defined in terms of the PT\cite{cor}
self-energy, obey Eq.~(\ref{eq:gam}) with $b_{\rm PT}=b_\xi$, and
Eq.~(\ref{eq:mas}) with $c_{\rm PT}=c_\xi=0$,\cite{prl} while $d_{\rm PT}$ is
currently unknown.
In Fig.~\ref{f:thr}, the NNLO results for $\Gamma_i$ are plotted versus 
$m_i$ for the five cases considered in Eq.~(\ref{eq:gam}).
The down-most and middle solid curves depict the LO and NLO expansions,
respectively, which are common to the five cases.
The up-most solid curve corresponds to the NNLO expansion for $i=2,\xi$.
We note that $b_1$ is negative, while the other coefficients $b_i$ are
positive.
In particular, the NLO and NNLO corrections to $\Gamma_1(m_1)$ cancel at
$m_1=1.415$~TeV.

The comparison between the masses and widths in the various on-shell and
pole schemes is particularly simple in the HHA:
\begin{eqnarray}
\frac{m_i-m_2}{m_2}&=&c_i\left(\frac{Gm_2^2}{\pi}\right)^2
+{\cal O}(\lambda^3),
\nonumber\\
\frac{\Gamma_i-\Gamma_2}{\Gamma_2}&=&(b_i+3c_i-b_2)
\left(\frac{Gm_2^2}{\pi}\right)^2+{\cal O}(\lambda^3).
\end{eqnarray}
These expressions lead immediately to Eq.~(\ref{eq:mrxi}), identical results
for $M^{\rm PT}/m_2$ and $\Gamma^{\rm PT}/\Gamma_2$, and to
Eq.~(\ref{eq:muni}).
They also lead to
\begin{eqnarray}
\frac{m_3}{m_2}&=&1+\frac{9}{512}G^2m_2^4,\qquad
\frac{\Gamma_3}{\Gamma_2}=1-\frac{9}{512}G^2m_2^4,
\nonumber\\
\frac{m_1}{m_2}&=&1+\frac{9}{128}G^2m_2^4,\qquad
\frac{\Gamma_1}{\Gamma_2}=1+\frac{9}{128}G^2m_2^4.
\end{eqnarray}
For large $m_2$, these expressions approximate rather well the numerical
results from the full SM. For instance, from Eq.~(\ref{eq:muni}) we have
$(M_u-m_2)/m_2=9.9\%$ and $(\Gamma_u-\Gamma_2)/\Gamma_2=39.7\%$
for $m_2=800$~GeV, instead of 11\% and 44\%, respectively, from the full
theory.

In order to analyze the scheme dependence of the above relations and the 
convergence properties of the corresponding perturbative series, one possible
approach\cite{agh} is to expand the relevant physical quantities in terms of
different masses $m_i$.
We illustrate this procedure with $m_2$ and $\Gamma_2$, which are the physical 
quantities that parametrize the conventional Breit-Wigner resonance amplitude,
proportional to $(s-m_2^2+im_2\Gamma_2)^{-1}$.
The relation $\Gamma_2(m_2)$ can be obtained directly from Eq.~(\ref{eq:gam}) 
or, via Eqs.~(\ref{eq:gam}) and (\ref{eq:mas}), from the expansions
\begin{eqnarray}
m_2&=&m_i\left[1-c_i\left(\frac{Gm_i^2}{\pi}\right)^2
-d_i\left(\frac{Gm_i^2}{\pi}\right)^3\right],
\label{eq:mtwo}\\
\Gamma_2&=&\frac{3}{8}Gm_i^3\left[1+a\frac{Gm_i^2}{\pi}
+(b_2-3c_i)\left(\frac{Gm_i^2}{\pi}\right)^2\right].
\label{eq:gtwo}
\end{eqnarray}
In the $m_i$-expansion scheme, for given $m_2$, one evaluates $m_i$ from 
Eq.~(\ref{eq:mtwo}) and $\Gamma_2$ from Eq.~(\ref{eq:gtwo}).
As the calculation of $\Gamma_2(m_i)$ through ${\cal O}(\lambda^n)$ only
requires the knowledge of $m_2(m_i)$ through ${\cal O}(\lambda^{n-1})$ and 
there is no term linear in $\lambda$ in Eq.~(\ref{eq:mtwo}), in LO (NLO), we
set $m_i=m_2$ and keep the first contribution (first and second contributions)
in Eq.~(\ref{eq:gtwo}), while in NNLO we retain the first two terms in
Eq.~(\ref{eq:mtwo}) and the three terms in Eq.~(\ref{eq:gtwo}).
In this manner, $m_2$ and $\Gamma_2$ are expanded to the same order in 
$\lambda$ relative to their respective Born approximations. 
Using as criterion of convergence the range throughout which the NNLO
corrections are smaller in magnitude than the NLO ones at fixed $m_2$, we find 
that the domains of convergence for the $m_1$, $m_2$, $m_3$, $M_\xi$, and
$M_u$ expansions are $m_2<733$~GeV, 930~GeV, 843~GeV, 930~GeV, and 672~GeV,
respectively.
In this connection, NLO (NNLO) correction means the difference between NLO and 
LO (NNLO and NLO) calculations.
We also find that these expansions, when restricted to overlapping domains of 
convergence, are in good agreement with each other.
Thus, the scheme dependence of the $\Gamma_2(m_2)$ relation is quite small
over the convergence domains of the expansions.

Another criterion that can be applied to judge the relative merits of the 
expansions is the closeness of the corresponding masses $m_i$ to $\bar m$, the 
peak position of the modulus of the $J=0$, iso-scalar Goldstone-boson
scattering amplitude.
The relation between $\bar m$ and $m_3$ is given to NNLO in the
literature.\cite{swi}
Using Eq.~(\ref{eq:mas}), we can get the corresponding expressions for
$i=1,2,\xi$.
In the case of $i=2$, we have
\begin{equation}
m_2=\bar m\left[1+\frac{3\pi^2}{64}\left(\ln2-\frac{5}{2}\right)
\left(\frac{G\bar m^2}{\pi}\right)^2-0.778\left(\frac{G\bar m^2}{\pi}\right)^3
\right].
\end{equation}
For $\bar m=800$~GeV, we find
$m_1=0.984\,\bar m$, $m_2=0.925\,\bar m$, $m_3=0.940\,\bar m$, and
$M_\xi=0.934\,\bar m$, while, for $\bar m=1$~TeV, we have
$m_1=0.954\,\bar m$, $m_2=0.797\,\bar m$, $m_3=0.836\,\bar m$, and
$M_\xi=0.829\,\bar m$.

\section{Conclusions for Second Subject}

i) In the SM, for a heavy Higgs boson, the differences between the
on-shell mass and width and their pole counterparts $(m_2,\Gamma_2)$ are
sensitive functions of the gauge parameter.
They become numerically large over a class of gauges that includes the unitary
gauge (about 11\% in the mass and 44\% in the width for $m_2=800$~GeV).
These features were overlooked in the extensive literature on the Higgs boson.
ii) For a light Higgs boson ($M_H\approx200$~GeV), the differences remain
small in all gauges, except in the vicinity of unphysical thresholds, where
our expansion is not valid.
iii) In the intermediate case ($M_H\approx400$~GeV), the differences reach
0.6\% in the mass and 3.3\% in the width.
iv) The PT mass and width remain close to $(m_2,\Gamma_2)$, reaching
differences of 0.7\% and $-0.7\%$, respectively, at $m_2=800$~GeV.
v) The differences of $M$, $M^{\rm PT}$, $\Gamma$, $\Gamma^{\rm PT}$ with
$(m_1,\Gamma_1)$ are somewhat larger.
vi) Theoretical expressions that relate the widths to the masses in NNLO
are available in the HHA $(g\to0$, $M_W\to0$, $\lambda\propto g^2M_H^2/M_W^2$
fixed). This is
precisely the order of expansion in which the gauge dependence sets in.
vii) In the HHA, the gauge dependence is reduced to a two-valued
expression, with one value corresponding to finite $\xi_W$, $\xi_Z$ ($R_\xi$
gauges), and another one corresponding to the unitary gauge.
The theoretical expressions for the mass and width differences become very
simple, and approximate rather well those obtained, for a heavy Higgs, in the
full SM.
viii) Using a convergence criterion based on the $(m_2,\Gamma_2)$ relation,
the $m_i$-expansions have domains of convergence with upper bounds
$(m_2)_{\rm max}$ in the range 672~GeV${}<(m_2)_{\rm max}<930$~GeV,
depending on what masses are employed.
The scheme dependence in the $(m_2,\Gamma_2)$ relation is small (${}\alt3\%$)
over overlapping domains of convergence, but not beyond.
Another criterion to judge the merits of the various expansion schemes is the
relative proximity of the corresponding masses to the peak energy.
ix) The fundamental importance of expansions based on the complex-pole
parameters is that they involve gauge-invariant quantities, namely masses
and widths that can be identified with physical quantities.

\section*{Acknowledgments}
The author is grateful to B. Kniehl and M. Passera for their collaboration.
He would also like to thank the members of the Max-Planck-Institut f\"ur
Physik for their warm hospitality, and the Alexander-von-Humboldt
Foundation for its kind support. This research was supported in part by
NSF grant No.\ PHY--9722083.

\end{document}